\documentclass[preprint,showpacs,preprintnumbers,amsmath,amssymb]{revtex4-1}
\usepackage{mathrsfs}
\usepackage{graphicx}
\usepackage{dcolumn}% Align table columns on decimal point
\usepackage{bm}% bold math
\begin{document}

\title{On angular momentum of photons: the role of transversality condition in quantum mechanics}

\author{Chun-Fang Li\footnote{Email address: cfli@shu.edu.cn}}

\affiliation{Department of Physics, Shanghai University, 99 Shangda Road, 200444
Shanghai, China}

\date{\today}% It is always \today, today,

\begin{abstract}

Whether the total angular momentum of the photon can be separated into spin and orbital parts has been a long-standing problem due to the constraint of transversality condition on its vector wavefunction. A careful analysis shows that the situation arises from a misuse of the constraint in quantum mechanics.
To use the constraint properly, we convert the vector representation into a two-component representation in which the wavefunction is not constrained by any conditions.
Upon doing so, we not only separate the spin conceptually from the orbital angular momentum but also identify the Berry gauge in which the two-component wavefunction can be canonically quantized. 
A corollary is that only in one particular Berry gauge can a radiation field be canonically quantized in terms of the plane waves.
The degree of freedom to fix the Berry gauge, which reflects a symmetry of the constraint, turns out to have observable quantum effects.

\end{abstract}

%\pacs{42.50.Tx, 03.65.Ca, 42.90.+m}% PACS, the Physics and Astronomy
%\keywords{}                       % Use showkeys class option if keyword
                                   % display desired
\maketitle

%----------------------------------------------------------------

\section{Introduction}

Even though the photon is a well-known boson of spin 1, it is still unclear how to separate its spin from its orbital angular momentum (OAM) in quantum mechanics \cite{Akhiezer, Jauch, Cohen, Barnett-A, EN, Enk, Barnett, Leader, Li-WY}.
As is known, the nonlocality of the photon in position space \cite{Jauch-P, Amrein, Pauli, Rosewarne-S} makes it impossible to define a commutative position operator \cite{Cohen, Pryce, Newton, Wightman, Jordan, Hawton} and to introduce a position-space wavefunction \cite{Akhiezer, Bialynicki1996, Hawton99-1} for the photon in the usual sense \cite{Sakurai}.
One can have only the momentum-space ($\mathbf k$-space) wavefunction \cite{Akhiezer, Cohen, Bialynicki1996}, a vector function $\mathbf{f}(\mathbf{k}, t)$ with $\mathbf k$ the wavevector. It satisfies the following Schr\"{o}dinger equation,
\begin{equation}\label{SE-V}
    i \frac{\partial \mathbf{f}}{\partial t}= \omega \mathbf{f},
\end{equation}
where the angular frequency $\omega =ck$ plays the role of Hamiltonian and $k=|\mathbf{k}|$.
Here there is one peculiar feature that does not usually occur in quantum mechanics. This is that the vector wavefunction is constrained by the transversality condition
\begin{equation}\label{TC}
    \mathbf{f}^\dag \mathbf{w}=0,
\end{equation}
where $\mathbf{w}=\mathbf{k}/k$ is the unit wavevector, the superscript $\dag$ stands for the conjugate transpose, and vectors of three components, such as $\mathbf f$ and $\mathbf k$, are treated as column matrices so that their scalar products are expressed as matrix multiplications. In other words, the vector representation is a constrained one.
In this representation, the operator for the spin is given by \cite{Akhiezer, Cohen}
\begin{equation}\label{S}
    \hat{\mathbf S}= \hbar \hat{\mathbf \Sigma},
\end{equation}
where
$(\hat{\Sigma}_k)_{ij} =-i \epsilon_{ijk}$ with $\epsilon_{ijk}$ the Levi-Civit\'{a} pseudotensor.
And the operator for the OAM about the coordinate origin of the laboratory reference system is given by
\begin{equation}\label{L}
    \hat{\mathbf L}= -\hat{\mathbf P} \times \hat{\mathbf X},
\end{equation}
where
\begin{subequations}\label{XandP}
\begin{align}
  \hat{\mathbf P} &= \hbar \mathbf{k},    \label{P}\\
  \hat{\mathbf X} &= i \nabla_\mathbf{k}, \label{X}
\end{align}
\end{subequations}
represent the momentum and position, respectively, and $\nabla_{\mathbf k}$ is the gradient operator with respect to $\mathbf k$.
The problem is that such an explicit separation between the spin and OAM has not yet been accepted as physically meaningful \cite{Akhiezer, Jauch, Cohen, Barnett-A, EN, Enk, Barnett, Leader}, due mainly to a faulty assumption on the role of the operators $\hat{\mathbf S}$ and $\hat{\mathbf L}$ in the constrained representation. A representative argument \cite{Cohen, EN, Enk} is as follows.
\begin{quote}
    The spin operator $\hat{\mathbf S}$ rotates the vector wavefunction $\mathbf f$ without rotating its variable $\mathbf k$; the OAM operator $\hat{\mathbf L}$ rotates $\mathbf k$ without rotating $\mathbf f$. But neither operation can preserve the transversality condition (\ref{TC}). So the separation of the spin from the OAM has no physical meaning.
\end{quote}

Obviously, the key point in that argument is the assumption that the operators $\hat{\mathbf S}$ and $\hat{\mathbf L}$ generate rotations in ordinary space.
However, such an assumption means \cite{EN, Enk, Bliokh} that they are both canonical variables, satisfying the canonical commutation relation of the angular momentum \cite{Sakurai},
\begin{subequations}
\begin{align}
  [\hat{S}_i, \hat{S}_j] & =i \hbar \epsilon_{ijk} \hat{S}_k,  \label{CCR-S}  \\
  [\hat{L}_i, \hat{L}_j] & =i \hbar \epsilon_{ijk} \hat{L}_k.  \label{CCR-L}
\end{align}
\end{subequations}
This observation indicates that the constraint of transversality condition (\ref{TC}) on the vector wavefunction does not necessarily rule out the separation of the spin from the OAM. Instead, it may rule out that the operators $\hat{\mathbf S}$ and $\hat{\mathbf L}$ satisfy the canonical commutation relation.
Indeed, if the OAM satisfies the canonical commutation relation (\ref{CCR-L}), it is required \cite{Sakurai} that the momentum and position are canonically conjugate to each other \cite{Cohen}, satisfying the following canonical commutation relations,
\begin{equation*}
[\hat{P}_i,\hat{P}_j] =0, \quad
[\hat{X}_i,\hat{X}_j] =0, \quad
[\hat{X}_i,\hat{P}_j] =i \hbar \delta_{ij}.
\end{equation*}
This is in direct contradiction with the nonlocality of the photon in position space, because the second equation means a commutative position operator $\hat{\mathbf X}$.
The purpose of this paper is to explore the role of the constraint (\ref{TC}) in quantum mechanics by investigating its effects on the properties of the spin and OAM.

If the constraint (\ref{TC}) were absent, the operators $\hat{\mathbf S}$ and $\hat{\mathbf L}$ would satisfy the commutation relations (\ref{CCR-S}) and (\ref{CCR-L}), respectively.
To investigate how the constraint (\ref{TC}) imposes its effects on the spin and OAM, we convert the vector representation into a two-component representation in which the wavefunction is free of any conditions.
By this it is meant that we introduce from the constraint (\ref{TC}) a quasi unitary matrix $\varpi$ the conjugate transpose of which maps the vector wavefunction onto a two-component wavefunction,
\begin{equation}\label{QUT-2}
    \tilde{f} =\varpi^\dag \mathbf{f}.
\end{equation}
In so doing we identify a degree of freedom that fixes the quasi unitary matrix and therefore the two-component representation. It appears as a Berry-gauge degree of freedom in the sense that it fixes a Berry-gauge potential \cite{Berry84}.
More importantly, the two-component wavefunction in a particular Berry gauge can be canonically quantized.
The canonical variables, either extrinsic or intrinsic, measure the properties of the photon in its ``own reference system'', in the language of Bertrand \cite{Bertrand}.
Because the Berry-gauge potential determines the ``own reference system'', the Berry-gauge degree of freedom is physically observable.
Let us explain the detail below.

\section{From constraint to two-component representation}\label{TC-representation}

As is well known \cite{Akhiezer, Cohen}, the Schr\"{o}dinger equation (\ref{SE-V}) together with the constraint (\ref{TC}) is equivalent to the free-space Maxwell's equations. The $\mathbf k$-space vector wavefunction uniquely determines the position-space electric and magnetic vectors that solve the free-space Maxwell's equations via
\begin{subequations}\label{E-and-H}
\begin{align}
  \mathbf{E} (\mathbf{X},t) & =\frac{1}{(2 \pi)^{3/2}}
                               \int \Big( \frac{\hbar \omega}{2 \varepsilon_0} \Big)^{1/2} \mathbf{f}
                                    \exp(i \mathbf{k} \cdot \mathbf{X}) d^3 k +c.c., \label{E-vec}\\
  \mathbf{H} (\mathbf{X},t) & =\frac{1}{(2 \pi)^{3/2}}
                               \int \Big( \frac{\hbar \omega}{2 \mu_0} \Big)^{1/2} \mathbf{w} \times \mathbf{f}
                                    \exp(i \mathbf{k} \cdot \mathbf{X}) d^3 k +c.c.,
\end{align}
\end{subequations}
respectively.
To explore its role in quantum mechanics, we will make use of the constraint (\ref{TC}) to introduce a representation in which the wavefunction is free of any constraints.

\subsection{From constraint to two-component wavefunction}

It is well known that the constraint (\ref{TC}) allows to expand the vector wavefunction in terms of a pair of orthogonal base vectors with respect to the wavevector. Let be $\mathbf u$ and $\mathbf v$ the two mutually perpendicular unit vectors that form with $\mathbf w$ a right-handed Cartesian triad
$\mathbf{uvw}$ \cite{Green, Mandel}, satisfying
\begin{equation}\label{triad}
    \mathbf{u} \times \mathbf{v} =\mathbf{w}, \quad
    \mathbf{v} \times \mathbf{w} =\mathbf{u}, \quad
    \mathbf{w} \times \mathbf{u} =\mathbf{v}.
\end{equation}
Choosing $\mathbf u$ and $\mathbf v$ as the basis as usual, we expand the vector wavefunction as
\begin{equation*}
    \mathbf{f} =\mathbf{u} f_\mathbf{u} +\mathbf{v} f_\mathbf{v},
\end{equation*}
where $f_\mathbf{u}=\mathbf{u}^\dag \mathbf{f}$ and $f_\mathbf{v}=\mathbf{v}^\dag \mathbf{f}$.
Putting the two coefficients together to introduce a two-component entity \cite{Li07-1, Li07-2}
$
    \tilde{f}
         =\bigg(
            \begin{array}{c}
              f_\mathbf{u} \\
              f_\mathbf{v} \\
            \end{array}
          \bigg),
$
we may rewrite it as
\begin{equation}\label{QUT-1}
    \mathbf{f} =\varpi \tilde{f},
\end{equation}
where $\varpi$ is a 3-by-2 matrix,
\begin{equation}\label{varpi}
\varpi=(
         \begin{array}{cc}
           \mathbf{u} & \mathbf{v} \\
         \end{array}
       ).
\end{equation}
Of course, we may choose any other two orthogonal unit vectors as the basis. To keep consistent with previous expressions \cite{Li07-1, Li07-2}, we adopt here the real-valued basis.

The matrix $\varpi$ in Eq. (\ref{QUT-1}) performs a quasi unitary transformation in the following sense.
On one hand, Eq. (\ref{QUT-1}) says that $\varpi$ acts on a two-component entity $\tilde f$ to give a vector wavefunction $\mathbf f$ that satisfies the constraint (\ref{TC}). It is easy to show that
\begin{equation}\label{unitarity-2}
    \varpi^{\dag} \varpi =I_2,
\end{equation}
where $I_2$ is the 2-by-2 unit matrix.
On the other hand, multiplying both sides of Eq. (\ref{QUT-1}) by $\varpi^{\dag}$ from the left and using Eq. (\ref{unitarity-2}), we get
\begin{equation*}
    \tilde{f} =\varpi^\dag \mathbf{f}.
\end{equation*}
This is Eq. (\ref{QUT-2}).
It says that $\varpi^{\dag}$ acts on a vector wavefunction to give a two-component entity. A straightforward calculation yields
$\varpi \varpi^{\dag} =I_3-\mathbf{w} \mathbf{w}^{\dag}$, where $I_3$ is the 3-by-3 unit matrix.
But when the constraint (\ref{TC}) is taken into account, we have
\begin{equation*}
    (\varpi \varpi^{\dag}) \mathbf{f}=\mathbf{f}.
\end{equation*}
Keeping in mind that $\varpi^{\dag}$ always acts on the vector wavefunction $\mathbf f$, we may rewrite it simply as
\begin{equation}\label{unitarity-1}
    \varpi \varpi^{\dag} =I_3.
\end{equation}
Eqs. (\ref{unitarity-2}) and (\ref{unitarity-1}) express the quasi unitarity \cite{Golub} of the transformation matrix $\varpi$. $\varpi^{\dag}$ is the Moore-Penrose pseudo inverse of $\varpi$, and vice versa.

From Eq. (\ref{unitarity-1}) it follows that
\begin{equation*}
    \tilde{f}^{\dag} \tilde{f} =\mathbf{f}^{\dag} \mathbf{f},
\end{equation*}
which means that the two-component entity shows up as another kind of $\mathbf k$-space wavefunction. The advantage of the two-component wavefunction over the vector wavefunction is that it is no longer subject to any constraints.
Multiplying both sides of Eq. (\ref{SE-V}) by $\varpi^{\dag}$ from the left and making use of Eqs. (\ref{QUT-2}) and (\ref{unitarity-1}), we arrive at the following Schr\"{o}dinger equation for the two-component wavefunction,
\begin{equation*}
    i \frac{\partial \tilde{f}}{\partial t}= \omega \tilde{f},
\end{equation*}
where the Hamiltonian is invariant under the quasi unitary transformation,
$\varpi^{\dag} \omega \varpi= \omega$.

\subsection{The degree of freedom to fix the two-component wavefunction}\label{introducingI}

It is noticed that Eqs. (\ref{triad}) cannot completely fix the quasi unitary matrix (\ref{varpi}), because they are not able to unambiguously fix the momentum-associated triad $\mathbf{uvw}$ up to a rotation about the wavevector \cite{Mandel}. To determine the two-component wavefunction for a given vector wavefunction, one has to figure out the way to fix the triad $\mathbf{uvw}$. Traditionally, this is done by introducing a constant real-valued unit vector, denoted here by $\mathbf I$, to define \cite{Stratton, Green, Pattanayak}
\begin{equation}\label{basis}
    \mathbf{u} (\mathbf I) =\mathbf{v} (\mathbf I) \times \frac{\mathbf k}{k},             \quad
    \mathbf{v} (\mathbf I) =\frac{\mathbf{I} \times \mathbf{k}} {|\mathbf{I} \times \mathbf{k}|}.
\end{equation}
Surprisingly, the unit vector $\mathbf I$ introduced this way turns out to be a degree of freedom \cite{Davis, Li2008, Li09-2, Li09-1, Wang, Yang, Li-Y} to fix the triad $\mathbf{uvw}$, because the unit vectors $\mathbf u$ and $\mathbf v$ defined in Eqs. (\ref{basis}) satisfy Eqs. (\ref{triad}) regardless of what the unit vector $\mathbf I$ is.
The quasi unitary matrix (\ref{varpi}) that follows from Eqs. (\ref{basis}) is thus a function of this degree of freedom,
\begin{equation*}
    \varpi=\varpi (\mathbf{I}).
\end{equation*}
Once the degree of freedom $\mathbf I$ is specified, the quasi unitary transformation (\ref{QUT-2}) expresses a one-to-one correspondence between the two-component wavefunction and the vector wavefunction. As a result, the two-component wavefunction following from Eq. (\ref{QUT-2}) constitutes a quantum-mechanical representation that is different from the vector representation.
Let us see how the wavefunction in the two-component representation depends on the degree of freedom $\mathbf I$ for a given vector wavefunction $\mathbf f$.

Suppose that the unit vector $\mathbf I$ is changed into a different one, $\mathbf{I}'$ say, so that the two-component wavefunction for the same vector wavefunction is given by
\begin{equation}\label{tildef'}
    \tilde{f}'=\varpi'^{\dag} \mathbf{f},
\end{equation}
where
$
\varpi'=(\begin{array}{cc}
            \mathbf{u}' & \mathbf{v}'
          \end{array}
        )
$
and
\begin{equation*}
    \mathbf{u}' =\mathbf{u} (\mathbf{I}')
                =\mathbf{v}' \times \frac{\mathbf k}{k}, \hspace{5pt}
    \mathbf{v}' =\mathbf{v} (\mathbf{I}')
                =\frac{\mathbf{I}' \times \mathbf{k}} {|\mathbf{I}' \times \mathbf{k}|}.
\end{equation*}
As remarked above, the transverse axes
$\mathbf{u}'$ and $\mathbf{v}'$
of the new triad
$\mathbf{u}' \mathbf{v}' \mathbf{w}$
are related to the transverse axes $\mathbf{u}$ and $\mathbf{v}$ of the old one by a rotation about $\mathbf k$.
Letting be $\phi(\mathbf k)$ the $\mathbf k$-dependent rotation angle, such a rotation can be expressed compactly by
\begin{equation}\label{pi-rotated}
    \varpi'=\exp [-i (\hat{\mathbf \Sigma} \cdot \mathbf{w}) \phi] \varpi,
\end{equation}
which can be rewritten as \cite{Tung}
\begin{equation}\label{rotation-pi}
    \varpi'=\varpi \exp \left(-i \hat{\sigma}_3 \phi \right)
\end{equation}
in terms of the Pauli matrix
\begin{equation*}
    \hat{\sigma}_3=\bigg(
                     \begin{array}{cc}
                       0 & -i \\
                       i &  0 \\
                     \end{array}
                   \bigg).
\end{equation*}
It is noted that the rotation angle $\phi$ in association with the change of the degree of freedom $\mathbf I$ is independent of the vector wavefunction.
Substituting Eq. (\ref{rotation-pi}) into Eq. (\ref{tildef'}) and making use of Eq. (\ref{QUT-2}), we find
\begin{equation}\label{GT-tildef}
    \tilde{f}'=\exp \left(i \hat{\sigma}_3 \phi \right) \tilde{f},
\end{equation}
showing that a given vector wavefunction can be expressed in terms of different two-component wavefunction in different two-component representation.

Eq. (\ref{GT-tildef}) is the transformation of the two-component wavefunction under the change of the degree of freedom $\mathbf I$.
When Eqs. (\ref{E-and-H}) are taken into account, this transformation is similar to the gauge transformation of the electromagnetic potentials in classical theory in the sense that it does not change the electric and magnetic vectors of the radiation field,
\begin{equation*}
    \varpi' \tilde{f}'=\varpi \tilde{f}= \mathbf{f}.
\end{equation*}
But it must be pointed out that the degree of freedom $\mathbf I$ is not the classical gauge degree of freedom at all. After all, the two-component wavefunction is not equivalent to the electromagnetic potentials.
As a matter of fact, it is the vector wavefunction that is equivalent to the vector potential in the Coulomb gauge \cite{Cohen}, because it has a one-to-one correspondence with that vector potential via
\begin{equation*}
    \mathbf{A} (\mathbf{X},t)=\frac{1}{(2 \pi)^{3/2} i}
    \int \Big( \frac{\hbar}{2 \varepsilon_0 \omega} \Big)^{1/2} \mathbf{f} \exp(i \mathbf{k} \cdot \mathbf{X}) d^3 k +c.c.
\end{equation*}
In view of this, we can say that the constraint (\ref{TC}) on the vector wavefunction is equivalent to the condition of Coulomb gauge,
$\nabla_\mathbf{X} \cdot \mathbf{A}=0$,
where $\nabla_\mathbf{X}$ is the gradient operator with respect to $\mathbf X$.

In a word, the constraint (\ref{TC}) makes the vector representation convertible into a two-component representation that is fixed by the degree of freedom $\mathbf I$. Now that the two-component wavefunction is free of any constraints, it is expected that the quantization of the radiation field in the two-component representation is a canonical one. In the remainder of this paper we will explore  the intrinsic and extrinsic canonical variables and their relations with the degree of freedom $\mathbf I$.

\section{The spin is commutative}\label{spin}

Let us first examine the spin. Its operator (\ref{S}) in the vector representation is transformed into
\begin{equation*}
    \hat{\mathbf s} =\varpi^{\dag} \hat{\mathbf S} \varpi
                    =\hbar \varpi^{\dag} \hat{\mathbf \Sigma} \varpi
\end{equation*}
in the two-component representation.
Upon decomposing the vector operator $\hat{\boldsymbol \Sigma}$ in the Cartesian triad $\mathbf{uvw}$ as
\begin{equation*}
    \hat{\boldsymbol \Sigma}
       =(\hat{\boldsymbol \Sigma} \cdot \mathbf{u}) \mathbf{u}
       +(\hat{\boldsymbol \Sigma} \cdot \mathbf{v}) \mathbf{v}
       +(\hat{\boldsymbol \Sigma} \cdot \mathbf{w}) \mathbf{w}
\end{equation*}
and taking Eqs. (\ref{triad}) into account, we get
\begin{equation}\label{s-sigma}
    \hat{\mathbf s}= \hbar \hat{\sigma}_3 \mathbf{w},
\end{equation}
where
\begin{equation*}
    \hat{\sigma}_3=\varpi^{\dag} (\hat{\mathbf \Sigma} \cdot \mathbf{w}) \varpi
\end{equation*}
is the Pauli matrix that we encountered in Eq. (\ref{rotation-pi}).
This shows that the spin of the photon lies entirely on its propagation direction. The Pauli matrix $\hat{\sigma}_3$ represents essentially the magnitude of the spin, known as the helicity.

From Eq. (\ref{s-sigma}) it immediately follows that the Cartesian components of the spin commute,
\begin{equation}\label{CR-s}
    [\hat{s}_i, \hat{s}_j]=0.
\end{equation}
This is what van Enk and Nienhuis \cite{EN, Enk} found in a second-quantization framework.
It is noted that the spin operator (\ref{s-sigma}) in the two-component representation is independent of the degree of freedom $\mathbf I$.

\section{Position operator and Berry-gauge degree of freedom}

In order to examine the OAM, we need consider the momentum and position separately.
It is seen from Eqs. (\ref{XandP}) that their operators in the two-component representation are given by
\begin{subequations}
\begin{align}
  \hat{\mathbf p} & =\varpi^{\dag} \hat{\mathbf P} \varpi
                    =\hbar \mathbf{k},                    \label{p} \\
  \hat{\mathbf x} & =\varpi^{\dag} \hat{\mathbf X} \varpi
                    =\hat{\boldsymbol \xi} +\hat{\mathbf b},          \label{x}
\end{align}
\end{subequations}
respectively, where
\begin{subequations}
\begin{align}
  \hat{\boldsymbol \xi} & =i \nabla_\mathbf{k},              \label{xi} \\
  \hat{\mathbf b}       & =i \varpi^{\dag} (\nabla_\mathbf{k} \varpi).   \label{bI}
\end{align}
\end{subequations}
Because the momentum operator (\ref{p}) is commutative,
\begin{equation}\label{CR-p}
    [\hat{p}_i, \hat{p}_j] =0,
\end{equation}
we are concerned mainly with the position operator (\ref{x}).

\subsection{Canonical position and ``own reference system''}

The position operator (\ref{x}) consists of two parts. The first part (\ref{xi}) is a gradient operator with respect to $\mathbf k$.
Because no constraints such as Eq. (\ref{TC}) exist for the two-component wavefunction, its Cartesian components commute,
\begin{equation}\label{CR-xi}
    [\hat{\xi}_i, \hat{\xi}_j] =0.
\end{equation}
Furthermore, it has the following commutation relation with the momentum operator,
\begin{equation}\label{CR-xip}
    [\hat{\xi}_i,\hat{p}_j] =i \hbar \delta_{ij}.
\end{equation}
The second part (\ref{bI}) is solely determined by the matrix $\varpi$. Straightforward calculations give
\begin{equation}\label{bI-sigma}
    \hat{\mathbf b}=\hat{\sigma}_3 \mathbf{A}_B,
\end{equation}
which is surprisingly proportional to the helicity, where
\begin{equation}\label{AI}
    \mathbf{A}_B=\frac{\mathbf{I} \cdot \mathbf{k}}{k |\mathbf{I} \times \mathbf{k}|} \mathbf{v}.
\end{equation}
Being commutative with the Hamiltonian, it represents a constant of motion. Moreover, its Cartesian components commute,
\begin{equation}\label{CR-b}
    [\hat{b}_i, \hat{b}_j]=0.
\end{equation}
With the help of Eqs. (\ref{CR-xi}) and (\ref{CR-b}), it is not difficult to find
\begin{equation*}
    \hat{\mathbf x} \times \hat{\mathbf x}
   =i \hat{\sigma}_3 \nabla_\mathbf{k} \times \mathbf{A}_B
   =i \hat{\sigma}_3 \mathbf{H}_B,
\end{equation*}
where
\begin{equation}\label{H}
    \mathbf{H}_B =\nabla_\mathbf{k} \times \mathbf{A}_B =-\frac{\mathbf w}{k^2}, \quad
    \mathbf{w} \neq \pm \mathbf{I}.
\end{equation}
The Cartesian components of the position do not commute. That is to say, the position of the photon is not canonically conjugate to the momentum.

Nevertheless, Eqs. (\ref{CR-p})-(\ref{CR-xip}) show that the first part represents such a position $\boldsymbol \xi$ that is canonically conjugate to the momentum, referred to as the canonical position.
It is of course different from the position of the photon in the laboratory reference system. According to Bertrand \cite{Bertrand}, it measures the position of the photon in its ``own reference system''.
Correspondingly, the second part denotes the own reference system for it represents the position vector from the coordinate origin of the laboratory reference system to the coordinate origin of the own reference system.
That the second part (\ref{bI}) is solely determined by $\varpi$ indicates that the own reference system is nothing but the $\mathbf I$-dependent triad $\mathbf{uvw}$. Let us explain this below.

\subsection{The two-component wavefunction is defined over the own reference system}

The gradient form (\ref{xi}) of the operator $\hat{\boldsymbol \xi}$ in the two-component representation demonstrates \cite{Sakurai} that the corresponding two-component wavefunction can always be viewed as the Fourier component of a function of the canonical position,
\begin{equation}\label{F-tilde}
    \tilde{F} (\boldsymbol{\xi}, t)
   =\frac{1}{(2\pi)^{3/2}} \int \tilde{f} (\mathbf{k}, t)
                                \exp(i\mathbf{k} \cdot \boldsymbol{\xi}) d^3 k,
\end{equation}
regardless of what the own reference system is. Such a conjugate relation can only be reasonably interpreted by the statement that the two-component wavefunction is defined over the own reference system. This is to be distinguished from the vector wavefunction.
As is known \cite{Akhiezer}, the gradient form (\ref{X}) of the operator $\hat{\mathbf X}$ in the vector representation demonstrates that the vector wavefunction can be viewed as the Fourier component of a function of $\mathbf X$, the position of the photon in the laboratory reference system,
\begin{equation}\label{F-vec}
    \mathbf{F}(\mathbf{X},t)
   =\frac{1}{(2 \pi)^{3/2}} \int \mathbf{f}(\mathbf{k}, t) \exp(i\mathbf{k} \cdot \mathbf{X}) d^3 k.
\end{equation}
This conjugate relation tells that the vector wavefunction is defined over the laboratory reference system.
In other words, the momentum in the two-component wavefunction should be distinguished quantum-mechanically from the momentum in the vector wavefunction, though they are quantitatively the same.
The former, conjugate canonically to the canonical position, is a canonical variable, referred to as the canonical momentum. It measures the momentum of the photon relative to the own reference system.
Whereas the latter, conjugate non-canonically to the position in the laboratory reference system, is not canonical. It measures the momentum of the photon relative to the laboratory reference system.

In fact, as can be seen from Eq. (\ref{QUT-2}), the components of $\tilde f$ in a particular two-component representation are the projections of $\mathbf f$ onto the base vectors $\mathbf u$ and $\mathbf v$. From the fact that these vectors themselves denote the transverse axes of the triad $\mathbf{uvw}$, it can be deduced that $\tilde f$ is defined over $\mathbf{uvw}$. That is to say, the own reference system is this $\mathbf I$-dependent triad.
Substituting Eq. (\ref{QUT-1}) into Eq. (\ref{F-vec}) and making use of Eq. (\ref{F-tilde}), we get
\begin{equation}\label{CI}
    \mathbf{F}(\mathbf{X},t)
   =\int \Pi(\mathbf{X}-\boldsymbol{\xi}) \tilde{F}(\boldsymbol{\xi},t) d^3 \xi,
\end{equation}
where
\begin{equation*}
    \Pi(\mathbf{X}-\boldsymbol{\xi})
   =\frac{1}{(2 \pi)^3} \int \varpi \exp[i \mathbf{k} \cdot (\mathbf{X}-\boldsymbol{\xi})] d^3 k.
\end{equation*}
According to Eq. (\ref{CR-xi}), the two-component function $\tilde{F} (\boldsymbol{\xi},t)$ can be interpreted as the probability amplitude of the position in the own reference system.
Nevertheless, because the relation (\ref{CI}) between $\mathbf{F}$ and $\tilde{F}$ is not a local one, the vector function $\mathbf{F}(\mathbf{X},t)$ cannot be interpreted as the probability amplitude of the position in the laboratory reference system.
We will see that the vector quantity (\ref{AI}) to determine the own reference system (\ref{bI-sigma}) is a Berry-gauge potential and the unit vector $\mathbf I$ is the degree of freedom to fix the Berry gauge.

\subsection{Unit vector $\mathbf I$ is a Berry-gauge degree of freedom}

Substituting Eq. (\ref{bI-sigma}) into Eq. (\ref{x}), we have
\begin{equation*}
    \hat{\mathbf x}=\hat{\boldsymbol \xi} +\hat{\sigma}_3 \mathbf{A}_B.
\end{equation*}
According to Barut and Bracken \cite{Barut}, if $\hat{\mathbf x}$, the position of the photon in the laboratory reference system, is regarded as the analog of the kinetic momentum of a charged particle in an external magnetic field and the canonical position $\hat{\boldsymbol \xi}$ is regarded as the analog of the canonical momentum, then $\hat{\sigma}_3$, the helicity of the photon, can be regarded as the analog of the electric charge of the particle and the vector quantity $\mathbf{A}_B$ can be regarded as the analog of the vector potential of the magnetic field.
That is, $\mathbf{A}_B$ appears as the gauge potential of the ``external'' gauge field (\ref{H}) on the helicity.

In a two-component representation that is indicated by a different unit vector, $\mathbf{I}'$, the operator for the position in the laboratory reference system is given by
\begin{equation*}
    \hat{\mathbf x}' =\varpi'^{\dag} \hat{\mathbf X} \varpi'
                     =\hat{\boldsymbol \xi} +\hat{\mathbf b}',
\end{equation*}
where $\varpi'$ is given by Eq. (\ref{rotation-pi}),
\begin{equation}\label{bI'}
    \hat{\mathbf b}'=\hat{\sigma}_3 \mathbf{A}'_B,
\end{equation}
and
\begin{equation*}
    \mathbf{A}'_B=\frac{\mathbf{I}' \cdot \mathbf{k}}{k |\mathbf{I}' \times \mathbf{k}|}
                  \mathbf{v}'.
\end{equation*}
Besides, from Eq. (\ref{bI}) it follows that
$\hat{\mathbf b}'=i \varpi'^\dag(\nabla_\mathbf{k} \varpi')$.
With the help of Eqs. (\ref{rotation-pi}) and (\ref{unitarity-2}), we get
\begin{equation*}
    \hat{\mathbf b}'
   =\hat{\mathbf b}+\hat{\sigma}_3 \nabla_{\mathbf k} \phi.
\end{equation*}
An inspection of Eqs. (\ref{bI'}) and (\ref{bI-sigma}) finds
\begin{equation}\label{GT-A}
    \mathbf{A}'_B =\mathbf{A}_B+\nabla_{\mathbf k} \phi,
\end{equation}
showing that the vector potential undergoes a ``gauge transformation'' under the change of the unit vector $\mathbf I$, with $\phi$ the corresponding gauge function.
It is thus concluded that $\mathbf I$ is the degree of freedom to fix the gauge potential.

However, the ``gauge transformation'' (\ref{GT-A}) is different from that of the vector potential of the magnetic field on a charged particle. This is because the gauge potential (\ref{AI}) has a one-to-one correspondence with the ``external'' field (\ref{H}), a Berry-gauge field \cite{Berry84} that corresponds to a ``magnetic monopole'' \cite{Dirac31} of unit strength in $\mathbf k$-space \cite{Bialynicki1987, Fang}. The gauge degree of freedom $\mathbf I$, called the Berry-gauge degree of freedom, denotes the ``location'' of the monopole's singular line. The ``gauge transformation'' (\ref{GT-A}), called the Berry-gauge transformation, plays the role of changing the location of the singular line so that
$\nabla_\mathbf{k} \times \nabla_\mathbf{k} \phi \neq 0$
when $\mathbf{w}=\pm \mathbf{I}$ or $\mathbf{w}=\pm \mathbf{I}'$.
Different Berry gauge indicated by different value of $\mathbf I$ means different ``external'' field.
The two-component representation is thus a Berry-gauge representation.

In analogy with the gauge transformation of the first kind on the wavefunction of a charged particle in a magnetic field \cite{Sakurai, Barut}, the above mentioned Eq. (\ref{GT-tildef}) is the Berry-gauge transformation of the first kind on the two-component wavefunction that corresponds to the potential transformation (\ref{GT-A}).
Specifically, analogous to the gauge invariance of the kinetic momentum of the charged particle, the position of the photon in the laboratory reference system is invariant under the Berry-gauge transformation,
\begin{equation*}
    \tilde{f}'^\dag \hat{\mathbf x}' \tilde{f}'=\tilde{f}^\dag \hat{\mathbf x} \tilde{f}.
\end{equation*}
Of course, the canonical momentum is invariant under the Berry-gauge transformation.

\section{The OAM does not generate spatial rotations}\label{OAM}

Now we are in a position to examine the OAM.
The same as the position operator (\ref{x}), the operator in the two-component representation for the OAM about the coordinate origin of the laboratory reference system also splits into two parts,
\begin{equation}\label{l}
    \hat{\mathbf l}= \varpi^\dag \hat{\mathbf L} \varpi
                   =-\hat{\mathbf p} \times \hat{\mathbf x}
                   = \hat{\boldsymbol \lambda} +\hat{\mathbf m}.
\end{equation}
The first part
\begin{equation*}
    \hat{\boldsymbol \lambda}=-\hat{\mathbf p} \times \hat{\boldsymbol \xi}
\end{equation*}
represents the OAM of the photon about the coordinate origin of the own reference system.
Thanks to the canonical commutation relations (\ref{CR-p})-(\ref{CR-xip}), it satisfies the canonical commutation relation of the angular momentum,
\begin{equation}\label{CR-lambda}
    [\hat{\lambda}_i, \hat{\lambda}_j] =i \hbar \epsilon_{ijk} \hat{\lambda}_k.
\end{equation}
For clarity, we will refer to it as the canonical OAM. Clearly, it is a constant of motion,
\begin{equation}\label{CR-lambda and omega}
    [\hat{\boldsymbol \lambda}, \omega] =0.
\end{equation}
The second part
\begin{equation}\label{mI}
    \hat{\mathbf m}
   \equiv \hat{\mathbf b} \times \hat{\mathbf p}
   =\hbar \hat{\sigma}_3 \frac{\mathbf{I} \cdot \mathbf{k}}{|\mathbf{I} \times \mathbf{k}|} \mathbf{u}
\end{equation}
represents the OAM of the photon concentrated at the coordinate origin of the own reference system. It depends on the helicity.
Different from $\hat{\boldsymbol \lambda}$, its Cartesian components commute,
\begin{equation}\label{CR-m}
    [\hat{m}_i, \hat{m}_j]=0.
\end{equation}
But like $\hat{\boldsymbol \lambda}$, it is a constant of motion,
\begin{equation}\label{CR-m and omega}
    [\hat{\mathbf m}, \omega] =0.
\end{equation}
According to Eqs. (\ref{CR-lambda and omega}) and (\ref{CR-m and omega}), the total OAM is a constant of motion, too.
It is noted that the expression (\ref{l}) for the total OAM has its counterpart in classical mechanics \cite{Goldstein}: the angular momentum of a system about a reference point is the angular momentum of the system concentrated at the barycenter plus the angular momentum of the system about the barycenter.
What is interesting here is that the total OAM is dependent on the Berry-gauge degree of freedom when expressed in terms of the canonical OAM.

The first part of the OAM does not commute with the spin,
\begin{equation*}
    [\hat{\lambda}_i, \hat{s}_j]= i \hbar \epsilon_{ijk} \hat{s}_k,
\end{equation*}
though the second part does. As a result, the total OAM does not commute with the spin,
\begin{equation}\label{CR-lands}
    [\hat{l}_i, \hat{s}_j]= i \hbar \epsilon_{ijk} \hat{s}_k.
\end{equation}
With the help of Eqs. (\ref{CR-lambda}) and (\ref{CR-m}), it is not difficult to find
\begin{equation}\label{CR-OAM}
    [\hat{l}_i, \hat{l}_j]
   =i \hbar \epsilon_{ijk} (\hat{l}_k-\hat{s}_k).
\end{equation}
This is the commutation relation of the OAM that was found by van Enk and Nienhuis in a second-quantization framework \cite{EN, Enk}.
It clearly shows that the OAM is not the generator of spatial rotations as is usually assumed \cite{Cohen, Bliokh}.

From Eqs. (\ref{s-sigma}), (\ref{l}), and (\ref{mI}) it follows that the operator for the total angular momentum in the two-component representation reads
\begin{equation*}
    \hat{\mathbf j}
   =\hat{\mathbf s} +\hat{\mathbf l}
   =\hat{\boldsymbol \lambda} +\hbar \hat{\sigma}_3 \frac{\mathbf{I} \times \mathbf{v}}
                                                         {\mathbf{I} \cdot  \mathbf{u}}.
\end{equation*}
It has a very interesting property that the component of $\hat{\mathbf j}$ along $\mathbf I$ is equal to the component of $\hat{\boldsymbol \lambda}$ along the same direction:
$\hat{\mathbf j} \cdot \mathbf{I}=\hat{\boldsymbol \lambda} \cdot \mathbf{I}$.
Note that the first part on the righthand side is not the orbital part of the total angular momentum though it is helicity-independent. Likewise, the second part is not the spin part though it is helicity-dependent.
This explains why the total angular momentum of a non-paraxial beam cannot be separated into helicity-independent OAM and helicity-dependent spin \cite{Barnett-A}.
With the help of Eqs. (\ref{CR-s}), (\ref{CR-lands}), and (\ref{CR-OAM}), it is easy to find
\begin{equation*}
    [\hat{j}_i, \hat{j}_j]= i \hbar \epsilon_{ijk} \hat{j}_k,
\end{equation*}
as is expected.

\section{Canonical quantization in a Berry gauge}

We have seen that the position and momentum of the photon in the own reference system are canonically conjugate to each other. Accordingly, the OAM of the photon about the coordinate origin of the own reference system is also canonical.
But unfortunately, the canonical OAM does not commute with the spin. This is a phenomenon that sheds new light on the intrinsic canonical variable of the photon.

\subsection{The intrinsic canonical variable in a Berry gauge}\label{meaning}

\subsubsection{Polarization is relative to the own reference system}

To explore the intrinsic canonical variable, we consider the Poincar\'{e} vector that is defined, according to Merzbacher \cite{Merz}, as follows,
\begin{equation*}
    \boldsymbol{\varsigma}
   =\frac{\tilde{f}^\dag \hat{\boldsymbol \sigma} \tilde{f}}{\tilde{f}^\dag \tilde{f}}
   =\varsigma_1 \mathbf{u}+\varsigma_2 \mathbf{v}+\varsigma_3 \mathbf{w},
\end{equation*}
where
\begin{equation}\label{v-sigma}
    \hat{\boldsymbol \sigma}
   =\hat{\sigma}_1 \mathbf{u} +\hat{\sigma}_2 \mathbf{v} +\hat{\sigma}_3 \mathbf{w},
\end{equation}
and $\hat{\sigma}_i$'s are the Pauli matrices,
\begin{equation}\label{PM}
    \hat{\sigma}_1=\bigg(\begin{array}{cc}
                           1 &  0 \\
                           0 & -1
                         \end{array}
                   \bigg), \quad
    \hat{\sigma}_2=\bigg(\begin{array}{cc}
                           0 & 1 \\
                           1 & 0
                         \end{array}
                   \bigg), \quad
    \hat{\sigma}_3=\bigg(\begin{array}{cc}
                           0 & -i \\
                           i &  0
                         \end{array}
                   \bigg).
\end{equation}
The role of the Poincar\'{e} vector is to characterize the polarization of the photon. In the case of a plane-wave state, its Cartesian components $\varsigma_i$ are the well-known Stokes parameters.
So the Pauli vector (\ref{v-sigma}) is the polarization operator in the two-component representation \cite{Jauch}.

But it is important to note that the polarization represented by the Pauli vector (\ref{v-sigma}) is relative to the own reference system, the momentum-associated triad $\mathbf{uvw}$, in the sense that the Pauli matrices represent the components of the polarization along the Cartesian axes of $\mathbf{uvw}$.
This is in consistency with the observation that the two-component wavefunction is defined over $\mathbf{uvw}$.
Because the Pauli matrices commute with the extrinsic canonical variables such as the canonical position and canonical momentum, it seems that only such a polarization can be regarded as the intrinsic degree of freedom.
More importantly, the Pauli matrices satisfy the canonical commutation relation of the angular momentum,
\begin{equation}\label{CR-sigma}
    [\hat{\sigma}_i, \hat{\sigma}_j]=2i \epsilon_{ijk} \hat{\sigma}_k,
\end{equation}
except for a factor two. In other words, so identified intrinsic degree of freedom is a canonical variable.

\subsubsection{Physical meaning of the polarization}

We have seen that due to the constraint (\ref{TC}), the vector wavefunction can be expressed in terms of a two-component wavefunction in one particular Berry gauge.
Because the vector wavefunction uniquely determines the electric and magnetic vectors via Eqs. (\ref{E-and-H}), the polarization in a Berry gauge is the canonical variable to describe the vectorial nature of the photon.
It should be pointed out that the same as the canonical position, the polarization is also Berry-gauge dependent.

In a different Berry gauge $\mathbf{I}'$, the Poincar\'{e} vector for the same vector wavefunction is given by
\begin{equation*}
    \boldsymbol{\varsigma}'
   =\frac{\tilde{f}'^\dag \hat{\boldsymbol \sigma}' \tilde{f}'}{\tilde{f}'^\dag \tilde{f}'}
   =\varsigma'_1 \mathbf{u}'+ \varsigma'_2 \mathbf{v}' +\varsigma'_3 \mathbf{w},
\end{equation*}
where
$\hat{\boldsymbol \sigma}'
=\hat{\sigma}_1 \mathbf{u}' +\hat{\sigma}_2 \mathbf{v}' +\hat{\sigma}_3 \mathbf{w}
$,
and $\tilde{f}'$ is given by Eq. (\ref{GT-tildef}). Straightforward calculations yield
\begin{eqnarray*}
  \varsigma'_1 &=&  \varsigma_1 \cos 2\phi +\varsigma_2 \sin 2\phi, \\
  \varsigma'_2 &=& -\varsigma_1 \sin 2\phi +\varsigma_2 \cos 2\phi, \\
  \varsigma'_3 &=& \varsigma_3.
\end{eqnarray*}
Because we have according to Eq. (\ref{rotation-pi}),
\begin{equation*}
    \mathbf{u}' = \mathbf{u} \cos \phi +\mathbf{v} \sin \phi, \quad
    \mathbf{v}' =-\mathbf{u} \sin \phi +\mathbf{v} \cos \phi,
\end{equation*}
it is not difficult to find
\begin{equation*}
    \boldsymbol{\varsigma}'=\varsigma_1 (\mathbf{u} \cos \phi-\mathbf{v} \sin \phi)
                           +\varsigma_2 (\mathbf{u} \sin \phi+\mathbf{v} \cos \phi)
                           +\varsigma_3 \mathbf{w},
\end{equation*}
which can be rewritten as
\begin{equation*}
    \boldsymbol{\varsigma}'=\exp[ i (\hat{\mathbf \Sigma} \cdot \mathbf{w}) \phi]
                            \boldsymbol{\varsigma}.
\end{equation*}
Apparently, it is different from $\boldsymbol \varsigma$.

\subsubsection{Comparison with the spin}

A comparison of Eq. (\ref{s-sigma}) with Eq. (\ref{v-sigma}) reveals an astonishing result that \emph{the spin is not the polarization}! It is only one part of the polarization, the longitudinal part. So the helicity, the magnitude of the spin, is the longitudinal component of the polarization.
This explains why the spin cannot ``give a complete description of the state of polarization'' \cite{Berry98}.
It is mentioned that because the polarization is Berry-gauge dependent, only in a particular Berry gauge can the helicity be considered as the longitudinal component of the polarization.

\subsection{Complete sets of canonical eigenfunctions in a Berry gauge}

It now becomes clear that the two-component representation in a particular Berry gauge is a canonical one.
There exist not only the extrinsic canonical variables such as the canonical position and canonical momentum but also the intrinsic canonical variable, the polarization.
So the wavefunction in the two-component representation can be canonically quantized in the usual way. That is, there exists in any two-component representation a complete set of canonical eigenfunctions.

It is well known that the canonical commutation relations (\ref{CR-p})-(\ref{CR-xip}) determine a complete set of three extrinsic quantum numbers that corresponds to a maximal set of three compatible extrinsic canonical variables. Moreover, the commutation relation (\ref{CR-sigma}) tells that the eigenvalue of any of the Pauli matrices (\ref{PM}) can be chosen as the intrinsic quantum number.
Choosing as usual the eigenvalue of the helicity as the intrinsic quantum number and denoting collectively by $q$ the set of extrinsic quantum numbers, we are given the following complete orthonormal set of canonical eigenfunctions,
\begin{equation}\label{EF}
    \tilde{f}_{\sigma q}=\tilde{\alpha}_\sigma f_q (\mathbf{k}) \exp(-i \omega t),
\end{equation}
where
\begin{equation*}
    \tilde{\alpha}_\sigma =\frac{1}{\sqrt{2}} \bigg(
                                                \begin{array}{c}
                                                      1 \\
                                                      i \sigma \\
                                                \end{array}
                                              \bigg)
\end{equation*}
is the eigenvector of the helicity operator $\hat{\sigma}_3$ with eigenvalue $\sigma= \pm 1$, satisfying
\begin{equation}\label{EV-eq}
    \hat{\sigma}_3 \tilde{\alpha}_{\sigma}= \sigma \tilde{\alpha}_{\sigma}.
\end{equation}
Here $f_q$ denotes the simultaneous normalized eigenfunction of the compatible extrinsic canonical variables, satisfying
\begin{equation*}
    \int f^\ast_{q'} f_q d^3 k=\delta_{q' q},
\end{equation*}
where the Kronecker $\delta_{q' q}$ should be replaced with the Dirac $\delta$-function for continuous extrinsic quantum numbers.

Nevertheless, it must be noted that the two-component eigenfunction (\ref{EF}), that is, the canonical quantum numbers $\sigma$ and $q$, does not yet determine an eigen state of the photon before the Berry-gauge degree of freedom is specified.
Here the Berry-gauge degree of freedom that fixes the matrix $\varpi$ determines, together with the quantum numbers $\sigma$ and $q$, the vector wavefunction of the eigen state as follows,
\begin{equation*}
    \mathbf{f}_{\sigma q}=\varpi \tilde{f}_{\sigma q},
\end{equation*}
which further determines the electric and magnetic vectors via Eqs. (\ref{E-and-H}).
At the same time, it determines through the Berry-gauge potential the own reference system of the eigen state in the sense that the expectation value of the coordinate origin of the own reference system is given by
\begin{equation*}
    \langle \hat{\mathbf b} \rangle_{\sigma q}
   =\sigma \frac{\int \mathbf{A}_B |f_q|^2 d^3 k}{\int |f_q|^2 d^3 k}.
\end{equation*}
It is pointed out that the Berry-gauge degree of freedom that van Enk and Nienhuis \cite{Enk} chose implicitly in their second-quantization framework is the unit vector $\mathbf{e}_Z$ along the $Z$ axis \cite{Wang, Yang}.

Consider an important case in which the compatible extrinsic canonical variables are chosen to be the Cartesian components of the canonical momentum. Denoting by $\hbar \mathbf{k}_0$ the eigen momentum, we have for the canonical eigenfunction,
\begin{equation}\label{PW}
    \tilde{f}_{\sigma \mathbf{k}_0}
   =\tilde{\alpha}_\sigma \delta^3 (\mathbf{k}-\mathbf{k}_0) \exp(-i \omega_0 t),
\end{equation}
where $\omega_0= c |\mathbf{k}_0|$. It is interesting to note that this is the eigenfunction of the operator $\hat{\mathbf b}$ in Eq. (\ref{bI-sigma}) with eigenvalue
\begin{equation}\label{EV-b}
    \mathbf{b}_{\sigma \mathbf{k}_0}
   =\frac{\sigma}{k_0}\frac{\mathbf{I} \cdot \mathbf{k}_0}{|\mathbf{I} \times \mathbf{k}_0|}
           \mathbf{v}_0,
\end{equation}
regardless of what the Berry gauge is, where
$\mathbf{v}_0=\frac{\mathbf{I} \times \mathbf{k}_0}{|\mathbf{I} \times \mathbf{k}_0|}$.
Substituting Eq. (\ref{PW}) into Eq. (\ref{F-tilde}) gives
\begin{equation*}
    \tilde{F}_{\sigma \mathbf{k}_0} (\boldsymbol{\xi},t)
   =\frac{\tilde{\alpha}_\sigma}{(2 \pi)^{3/2}}
    \exp{[i(\mathbf{k}_0 \cdot \boldsymbol{\xi}-\omega_0 t)]}.
\end{equation*}
It is nothing but a plane wave over the own reference system.

\section{Berry-gauge degree of freedom is physically observable}\label{observable}

Because the canonical variables, either extrinsic or intrinsic, describe only the physical properties of the photon in its own reference system, the Berry-gauge degree of freedom that determines the own reference system through the Berry-gauge potential (\ref{AI}) has observable physical effects.
To explain this, it is only necessary to prove that there exists a physical manipulation that changes the Berry-gauge degree of freedom of a photon state without changing its canonical quantum numbers, or equivalently, without changing its two-component wavefunction.

To this end, we notice that both the Schr\"{o}dinger equation (\ref{SE-V}) and the constraint (\ref{TC}) are invariant under the rotation of the vector wavefunction about the wavevector,
\begin{equation}\label{f-rotated}
    \mathbf{f}'=\exp [ -i (\hat{\mathbf \Sigma} \cdot \mathbf{w}) \phi ] \mathbf{f}.
\end{equation}
Substituting Eq. (\ref{QUT-1}) into Eq. (\ref{f-rotated}) and making use of Eq. (\ref{pi-rotated}), we have
\begin{equation}\label{QUT'}
    \mathbf{f}'=\varpi' \tilde{f}.
\end{equation}
It is a state that shares the same two-component wavefunction with the state (\ref{QUT-1}) but has a different Berry-gauge degree of freedom.
This demonstrates that Eq. (\ref{f-rotated}) describes exactly the physical manipulation mentioned above.
Let us show that such a change of the Berry-gauge degree of freedom is responsible for the so-called spin Hall effect of light~\cite{Onoda} in a refraction process \cite{Hosten}.

\subsection{Explanation of the spin Hall effect}

It was previously shown \cite{Li09-2} that in the first-order approximation, the refraction of a light beam at the interface between two dielectric media is such a physical process that changes the Berry-gauge degree of freedom of the beam with respect to the propagation direction.
To see how the change of the Berry-gauge degree of freedom is responsible for the spin Hall effect, let us confine ourselves to such a case in which the two-component wavefunction satisfies \cite{Yang}
\begin{equation}\label{mean-xi}
    \int \tilde{f}^\dag \hat{\boldsymbol \xi} \tilde{f} d^3 k=0.
\end{equation}
In this case, the coordinate origin of the own reference system reduces to the barycenter of the photon in the sense that
\begin{equation*}
    \langle \hat{\mathbf b} \rangle
   =\langle \hat{\mathbf x} \rangle
   =\frac{\int \tilde{f}^\dag \hat{\mathbf b} \tilde{f} d^3 k}{\int \tilde{f}^\dag \tilde{f} d^3 k}.
\end{equation*}
Furthermore, we confine ourselves to the eigenstate of the helicity,
\begin{equation}\label{Tf-sigma}
    \tilde{f}_\sigma=\tilde{\alpha}_\sigma f.
\end{equation}
The expectation value of the barycenter then becomes
\begin{equation*}
    \langle \hat{\mathbf b} \rangle_\sigma
   =\sigma \frac{\int \mathbf{A}_B |f|^2 d^3 k}{\int |f|^2 d^3 k}.
\end{equation*}

For a quasi plane wave that can be approximately described by the eigenfunction (\ref{PW}), if the angle between its Berry-gauge degree of freedom and $\mathbf{k}_0$ is much larger than its divergence angle, the Berry potential
$\mathbf{A}_B$
in the integrant of the numerator remains approximately the same within the region in which $|f|^2$ is appreciable.
In this case, the above expression becomes
\begin{equation*}
    \langle \hat{\mathbf b} \rangle_\sigma
   \approx \frac{\sigma}{k_0}\frac{\mathbf{I} \cdot \mathbf{k}_0}{|\mathbf{I} \times \mathbf{k}_0|}
           \mathbf{v}_0.
\end{equation*}
This is the same as Eq. (\ref{EV-b}).
As was shown in Ref. \cite{Li09-2}, in the experiments by Hosten and Kwiat \cite{Hosten}, the Berry-gauge degree of freedom of the incident wave satisfying Eq. (\ref{mean-xi}) is perpendicular to the propagation direction. So the barycenter of the incident wave lies on the propagation axis.
But the Berry-gauge degree of freedom of the refracted wave, which is parallel to the incidence plane, is no longer perpendicular to the propagation direction. Denoting by $\Theta$ the angle it makes with the propagation direction, the barycenter of the refracted wave is displaced transversely from the propagation axis by a distance
$(\sigma/k_0) \cot \Theta$. This is what Hosten and Kwiat observed \cite{Li09-2}.

\subsection{Connection with the Berry phase}

As is shown, the two vector wavefunctions (\ref{QUT-1}) and (\ref{QUT'}) that are different from each other only in the Berry-gauge degree of freedom represent different photon states.
Substituting Eq. (\ref{Tf-sigma}) into Eq. (\ref{QUT-1}), we have
\begin{equation*}
    \mathbf{f}_\sigma=\varpi \tilde{f}_\sigma.
\end{equation*}
Substituting Eq. (\ref{Tf-sigma}) into Eq. (\ref{QUT'}) and making use of Eqs. (\ref{rotation-pi}) and (\ref{EV-eq}), we find
\begin{equation*}
    \mathbf{f}'_\sigma=\exp(-i \sigma \phi) \mathbf{f}_\sigma,
\end{equation*}
which is different from the above vector wavefunction only by a helicity-dependent phase.
This reveals an unexpected result that the two vector wavefunctions, $\mathbf{f}'_\sigma$ and $\mathbf{f}_\sigma$, that are different from each other by a phase do not represent the same photon state.
It deserves emphasizing that such a phase does not result from the dynamical evolution governed by the Schr\"{o}dinger Eq. (\ref{SE-V}). Instead, it results from the change of the Berry-gauge degree of freedom and is therefore a Berry phase.
This is why the spin Hall effect of light that occurs in such a process \cite{Hosten} can be explained in terms of the Berry phase as well \cite{Onoda}.

\section{Conclusions and remarks}\label{remarks}

In summary, it is shown that the radiation field can be canonically quantized in one particular Berry gauge. The photon in free space seems not truly ``free'' when viewed from its own reference system. Its helicity is ``subject'' to a Berry-gauge field.
What deserves noting is that different from the spin of the nonrelativistic electron, here the intrinsic canonical variable represented by the Pauli matrices is the polarization of the photon relative to the momentum-associated triad $\mathbf{uvw}$. In addition, the factor two in the commutation relation (\ref{CR-sigma}) suggests that the operator (\ref{v-sigma}) be replaced with $\hat{\boldsymbol \sigma}/2$ if the polarization is stipulated to be a strict canonical variable that generates \cite{Jauch} rotations of the two-component wavefunction in the triad $\mathbf{uvw}$.

\section*{Acknowledgments}

The author is indebted to Vladimir Fedoseyev and Zihua Xin for their helpful discussions. This work was supported in part by the National Natural Science Foundation of China (60877055).

\end{document}